\documentclass[prd,twocolumn, showpacs,nofootinbib]{revtex4}

\usepackage{amstext,amsmath,amssymb}
\usepackage[dvips]{graphicx}
\usepackage{latexsym}

\newcommand{\Ref}[1]{(\ref{#1})}

\def\be{\begin{equation}}
\def\ee{\end{equation}}
\def\bes{\begin{eqnarray}}
\def\ees{\end{eqnarray}}

\def\f{\frac}

\def \vphi{\varphi}

\def\hh{{\cal H}}
\def\mn{{\mu\nu}}

\def\eps{\epsilon}
\def\ka{\kappa}

\newcommand{\lalg}[1]{\mathfrak{#1}}

\newcommand{\SO}{\mathrm{SO}}

\newcommand{\so}{\lalg{so}}

\def\eps{\epsilon}

\def\mn{{\mu \nu}}

\begin{document}

\title{Non-Commutativity of Effective Space-Time Coordinates and the Minimal Length}
%\title{DSR from quantum undeterminacy}

%\author{{\bf Florian Girelli}\footnote{girelli@sissa.it}, {\bf Etera R. Livine}\footnote{elivine@perimeterinstitute.ca}}
%\affiliation{SISSA and INFN, 4 via Beirut, Trieste, 34014, Italy}
%\affiliation{Laboratoire de Physique, ENS Lyon, CNRS UMR 5672, 46 All\'ee d'Italie, 69364 Lyon Cedex 07}

\author{{\bf Florian Girelli}\footnote{girelli@sissa.it}}
\affiliation{SISSA, 4 via Beirut, Trieste, 34014, Italy and INFN sezione de Trieste}
\author{{\bf Etera R. Livine}\footnote{elivine@perimeterinstitute.ca}}
\affiliation{Laboratoire de Physique, ENS Lyon, CNRS UMR 5672, 46 All\'ee d'Italie, 69364 Lyon Cedex 07}

\begin{abstract}
Considering that a position measurement can effectively involve a momentum-dependent shift and
rescaling of the ``true" space-time coordinates, we construct a set of effective space-time
coordinates which are naturally non-commutative. They lead to a minimum length and are shown to be
related to Snyder's coordinates and the five-dimensional formulation of Deformed Special
Relativity. This effective approach then provides a natural physical interpretation for both the
extra fifth dimension and the deformed momenta appearing in this context.

%Considering that a position measurement could induce effectively a shift and a rescaling momentum
%dependent, we construct a set of non-commutative effective space-time coordinates involving a
%minimum length, related to Snyder's coordinates and the five-dimensional formulation of Deformed
%Special Relativity. This effective approach provides then a natural physical interpretation for
%both  the extra fifth dimension  and the different momenta appearing in this context.
\end{abstract}

\maketitle

%%%%%%%%%%%%%%%%%%%%%%%%%%%%%%%%%%%%%%%%%%%5

%%%%%%%%%%
\section{The motivation: Implementing a minimal length}
%%%%%%%%%%
%
The goal of quantum gravity is to build a theory encompassing both quantum field theory and general
relativity. An expected feature is the existence of a minimal length scale defined by the Planck
length $l_P\,\equiv\,\sqrt{\hbar G/c^3}$. The usual issue is how to reconcile such a discrete
structure with the requirement of Lorentz invariance (or more generally diffeomorphism invariance).
We point out that such concepts of Lorentz invariant minimal length already exist in both general
relativity and quantum field theory as soon as we deal with massive objects.\\
In general relativity, a particle of mass $m$ creates a Schwarzschild metric with an event horizon
at the distance $r=l_S=2Gm/c^2$. This event horizon is a Lorentz invariant boundary: from the point
of view of a static observer (at infinity), the distance $r$ between the particle and a test
particle will get contracted under boosts but $r$ will always remain larger than $l_S$. The
curvature of space-time deforms the length contraction of special relativity and creates such a
bound\footnotemark.\\
\footnotetext{In fact, the Schwarzschild metric forbids to observe locally a real particle and thus contradicts
the standard formulation of the equivalence principle for real systems: the space-time is not flat
very close to a particle.}
In quantum field theory also, in presence of a massive field of mass $m$, the Compton length
$l_C=\hbar/mc$ establishes a minimal length scale. If one tries to probe a distance $r$ smaller
than $l_C$ then the vacuum fluctuations and the creation of virtual particles will blur the
measurement.\\
The issue is then to provide a unified framework and language to describe the same physical
phenomenon which seems due to two different causes in the two theories. Deformed (or doubly)
special relativity (DSR) and its %($\kappa$-Poincar\'e)
non-commutative geometry are such an attempt
\cite{DSR}. As we would expect in a quantum geometry theory, it defines the length/distance as a
quantum operator and the minimal length comes from a discrete spectrum of the operator (or more
generally a ``length" gap).\\
%
%\medskip
%
We propose to recover such a framework with non-commutative space-time coordinates assuming that
the space-time coordinates that we measure are effectively not the bare usual $x_\mu$ but objects
which also depend on the momentum $p_\mu$. The motivation behind this is that the mass (and
momentum) of a particle is fundamental to both the Schwarzschild radius and Compton length. Indeed,
on the one hand, the momentum deforms the space-time metric and will thus affect the measured
space-time coordinates; on the other hand, the momentum affects the position in quantum mechanics
due to the uncertainty principle.\\
%
%Finally, we also point out the relativity of the concept of mass itself. On the one hand, the
%gravitational interaction naturally creates a screening effect of the mass, so that the measured
%mass of an object depends on the scale of observation (or distance to the observer), e.g see the
%Brown-York mass \cite{BYmass}. On the other hand, quantum field theory allows fluctuations from the
%mass-shell\footnotemark, so $p^2$ does not necessarily give the mass $m^2$.
%
%\footnotetext{We refer to the renormalisation of the propagator in interacting field theories.
%We also keep in mind that the vacuum energy affects the mass/energy content of the region around
%any particle and thus changes the measured mass of that particle.}
%
Here, we introduce a class of momentum-dependent space-time coordinates. Requiring Lorentz covariant
coor\-dinates and focusing on the simplest examples, we ana\-lyze in details the cases of a
coordinate shift in $p_\mu$ and a $p^2$-dependent rescaling. The shift can be interpreted as a
dragging or time-lapse in the measurements, while the $p^2$-rescaling can be understood as the
effect on the measurements of the object's mass\footnotemark~deforming the surrounding space-time.
We show that these effective coordinates naturally lead to a stable $\so(4,1)$ (or $\so(3,2)$)
structure similar to deformed special relativity which represents a minimal length scale. This
shows how easy it is to get non-commutative space-time coordinates in special relativity. We then
relate this five-dimensional structure to the previous proposal of an extended special relativity
\cite{ESR}.\\
\footnotetext{
Since we work off-shell, the (squared) ``mass"  of the particle/object is defined as $p^2$ and is
not assumed to be constant. This allows our results to apply to quantum field theory, which allows
fluctuations off the mass-shell.}
We insist that we work within special relativity and with Lorentz covariant objects. We neither
break nor deform the Lorentz invariance. Finally, we call $\eta_\mn$ the flat space-time metric and
work with the signature $(+---)$.

%%%%%%%%%%
\section{Non-Commutativity of Effective Coordinates}
%%%%%%%%%%

A first possible effect is a dragging of the particle motion, that is we measure the position a bit
later (or earlier) than it actually is. This leads to introducing the following class of phase
space functions:
\be
X_\mu\,\equiv\, x_\mu-\f{\vphi}{\ka^2} p_\mu,
\ee
where $\vphi$ is a dimensionless Lorentz invariant function on the phase space and $\ka$ an
arbitrary mass scale here only for dimension purposes. Since we require $\vphi$ to be a scalar, it
can be a function of $x^2$, $p^2$ or the dilatation $D\,\equiv x_\nu p^\nu$. We do not inquire a
possible dependence on $x^2$ since we would like to focus on momentum-dependence\footnotemark. It
is also easy to check that function of $p^2$ does not change the Poisson brackets. This leaves the
case of a function $\vphi(D)$. The Poisson brackets are straightforward to compute:
\bes
\{X_\mu,X_\nu\}&=& -\f{\vphi'(D)}{\ka^2}j_\mn \nonumber\\
\{X_\mu,p_\nu\}&=& \eta_\mn -\vphi'(D)\f{p_\mu p_\nu}{\ka^2},
\label{class1}
\ees
where $j_\mn=x_\mu p_\nu -x_\nu p_\mu$ are the Lorentz generators.

\footnotetext{For a deformation of the type $\vphi(x^2)$, we compute:
$$
\{x_\mu+\vphi(x^2) p_\mu,x_\nu+\vphi(x^2) p_\nu\}=-(\vphi^2)'(x^2) j_\mn.
$$
The deformation generically depends on the distance $x^2$ but we get a constant deformation
parameter for $\vphi(x^2)=\sqrt{Ax^2+B}$. We obtain a similar structure of the brackets for the
class of effective coordinates defined by a $p^2$-dependent rescaling of the coordinates (cf Eq. \eqref{coordrescaled}).}

This new algebra of position-momentum is very similar to the algebra underlying deformed special
relativity. More precisely, if we require that the Poisson algebra \Ref{class1} closes, it means
that $\vphi'$  is constant i.e. $\vphi(D)$ linear in $D$. We neglect the constant term in $\vphi$
since a shift $\pm T p_\mu$ with constant $T$ amounts to a simple time shift on the trajectory.
Then up to a renormalisation of the mass scale $\ka$, we have two possibilities for $\alpha=\pm$:
\be
X_\mu \,=\, x_\mu+\alpha\f{D}{\ka^2} p_\mu.
\ee
This gives the commutators $\{X_\mu, X_\nu\}=-\alpha j_\mn/\ka^2$. And the $X,j$'s form a closed
Lie algebra, $\so(4,1)$ for $\alpha=+$ and $\so(3,2)$ for $\alpha=-$. This is exactly the structure
behind DSR: if we assume that we measure the coordinates $X_\mu$, thus that the $X$'s are more
physically relevant than the $x$'s, then we end up with non-commutative space-time coordinates of
the DSR type. It also means that there is a natural $\so(4,1)$ (and $\so(3,2)$) structure in
special relativity.

From the $\{X,X\}=\pm j/\ka^2$ commutation relations, the effective coordinates $X_\mu$ are
identified with five-dimensional Lorentz generators $j_{\mu 4}/\ka$, thus reproducing Snyder's
original proposal \cite{snyder}. At the quantum level, the eigenvalues of the $X_\mu$ are either
discrete or continuous depending on the 5d signature (discrete for space coordinates $X_i$ and
continuous for the time-like $X_0$ for $\so(4,1)$ and vice-versa). We can also compute the spectrum
of the space-time interval $X^2$ and of the spatial distance $X_iX_i$, which turn out to be
discrete in some cases. The interested reader will find more details in \cite{spectrum}.

Furthermore the coordinates $X_\mu=x_\mu-Dp_\mu/\ka^2$ are weak observables for the relativistic
particle of mass $m^2=\ka^2$, i.e. their Poisson bracket with the Hamiltonian constraint
$\hh\,\equiv p^2-\ka^2$ vanishes on the mass-shell. Thus there are also natural space-time
coordinates from this point of view. A detailed analysis of their relation to strong Dirac
observables and of the quantization of the relativistic particle in term of these coordinates can
be found in \cite{obs}.

\medskip

We now consider another class of effective coordinates defined by a rescaling of the space-time by
a momentum-dependent factor:
\be\label{coordrescaled}
X_\mu\,\equiv\,f\left(\f{p^2}{\kappa^2}\right)x_\mu
\ee
where $f$ is an arbitrary function and $\kappa$ still an arbitrary mass scale. On the mass-shell
when $p^2$ is held fixed, we can not distinguish measurements of the original coordinates $x$ and
of the modified coordinates $X$. However, as soon as $p^2$ is allowed to fluctuate, the behavior of
$X$ will differ from $x$.

%We would like to consider the $\tx$'s as effective coordinates taking into account the possibility
%of going off-shell in quantum field theory.

We can interpret this rescaling as a mass-dependence in the metric, similarly to what happens in
general relativity when massive objects deform the flat metric.

As above, these effective coordinates are also non-commutative:
\be
\{X_\mu,X_\nu\}\,=\,
-\f{(f^2)'}{\kappa^2}\,j_\mn,
%\f{-2}{\kappa^2}ff'\left(\f{p^2}{\kappa^2}\right)\,j_\mn,
\ee
where the argument $(p^2/\ka^2)$ is implicit. We introduce a dual rescaling of the momentum,
$P_\mu\,\equiv\,p_\mu/f$. The $\{X,P\}$ bracket then takes the same shape as above\footnotemark:
\be
\{X_\mu,P_\nu\}\,=\,
\eta_\mn-(f^2)'\f{P_\mu P_\nu}{\ka^2}.
\ee
The Lorentz generators are unmodified, $j_\mn=x_{[\mu}p_{\nu]}=X_{[\mu}P_{\nu]}$, and the new
coordinates $X_\mu$ and $P_\mu$ transform normally under Lorentz transformations.

\footnotetext{
More generally, if we introduce an arbitrarily rescaled momentum
$P_\mu\,=\,g(\f{p^2}{\ka^2})\,p_\mu$, we get the modified bracket:
$$
\{X_\mu,P_\nu\}\,=\,fg\,\left(\eta_\mn+\f{2g'}{g^3}\f{P_\mu P_\nu}{\ka^2}\right).
$$
To keep a leading order in $\eta_\mn$, it is natural to require that $fg=1$ and therefore
$P_\mu=p_\mu/f$. }

The $\{X,X\}\,\propto\,j$ commutation relation suggests an underlying five-dimensional structure.
As previously, as soon as the non-commutativity factor $(f^2)'$ is constant, the Poisson algebra
$(X_\lambda,j_\mn)$ closes and forms a $\so(4,1)$ or $\so(3,2)$ Lie algebra.
%When $f^2$ is not constant, we get an anomaly in the $\{(f^2)' j, X\}$ bracket and the Poisson
%algebra does not close:
%$$
%\{(f^2)' j_\mn, X_\lambda\}
%\,=\,
%(f^2)'\eta_{\lambda[\mu}X_{\nu]}-2f^2(f^2)''j_\mn P_\lambda.
%$$
This requirement means that the deformation must be of the type $f=\sqrt{Ap^2/\ka^2+B}$ with
arbitrary constants $A,B$. Up to a renormalisation of the mass scale $\ka$, we parameterize these
possible deformations as:
\be
f=\sqrt{\eps \f{p^2}{\ka^2}+\eps'}, \qquad \eps=\pm, \quad \eps'=\pm 1,0.
\ee
Notice that this can never be defined on the whole momentum space and we always get a minimal or
maximal bound on $p^2$ given by $\pm\ka^2$ or 0 depending on $\eps$ and $\eps'$. The sign $\eps=+$
gives a $\so(4,1)$ algebra while $\eps=-$ corresponds to a $\so(3,2)$ signature.

We can of course consider a generic deformation with a momentum dependent commutator $\{X,X\}$. For
example, it might be interesting to construct coordinates which would remain commutative for small
$p^2$ but lead to a constant non-commutativity for very large $p^2$, or vice-versa
(non-commutativity for small $p^2$ but classical in the asymptotic regime). However the algebra of
$X,j,p$ would not close.

\medskip

To sum up, assuming that the momentum has non-trivial effects on the measurement of space-time
coordinates and that we can model such effects by introducing effective coordinates $X$ depending
on both the original coordinates $x$ and the momentum $p$, we have shown that the commutativity of
the coordinates is not stable and that we naturally end up with non-commutative space-time
coordinates. Moreover, for Lorentz covariant effective coordinates, we get Poisson brackets
$\{X,X\}$ proportional to the Lorentz generators $j_\mn$, thus embedding our effective coordinates
in a five-dimensional structure with an underlying $\SO(4,1)$ (or $\SO(3,2)$) symmetry.

Next, we inquire in more details at this $\so(4,1)$ structure and relate it to the 5d
representation of DSR \cite{jerzy} and the recently proposed extended special relativity
\cite{ESR}.

%%%%
\section{The 5d Structure and Extended Special Relativity}
%%%%

Let us consider a more general possibility of both a shift and a rescaling of the space-time
coordinates:
\be
X_\mu\,\equiv\,f(\f{p^2}{\ka^2})\,\left[x_\mu-\vphi(D)\f{p_\mu}{\ka^2}\right].
\ee
We compute the position commutator:
\be
\{X_\mu,X_\nu\}\,=\,
-\f{j_\mn}{\ka^2}\,\left[(f^2)'+\vphi'\left(f^2-\f{p^2}{\ka^2}(f^2)'\right)\right],
\ee
where the arguments of $f$ and $\vphi$ are kept implicit. Once again, as soon as $(f^2)'$ and
$\vphi'$ are constant, the Poisson brackets define a closed Lie algebra $(X,j)$. Taking as above
$$
f^2=\eps\f{p^2}{\ka^2}+\eps', \quad \vphi=\alpha D,
$$
we obtain:
$$
\{X_\mu,X_\nu\}\,=\,
-\f{j_\mn}{\ka^2} (\eps+\eps'\alpha).
$$
The effective coordinates $X_\mu$ are thus identified with extended Lorentz generators $j_{\mu
4}/\ka$ and form a $\so(4,1)$ or $\so(3,2)$ algebra with the $j$'s depending on the values of the
parameters $\eps,\eps',\alpha$.

Such a structure reminds of the conformal group. The coordinates $X_\mu$ are actually very similar
to the generators $K_\mu$ of the special conformal transformations\footnotemark~but the main
difference is that $X$ depends on the arbitrary mass scale $\ka^2$ while the mass scale for $K$ is
set by the momentum $p^2$ itself.

\footnotetext{The generators  of the special conformal transformations in momentum space are:
$$
K_\mu\,\equiv\, p^2\left(x_\mu-2\f{D}{p^2}p_\mu\right).
$$
This is to be compared to $X_\mu=f(p^2)\,(x_\mu-\alpha D p_\mu/\ka^2)$. }

The identification of $X_\mu$ with the generators $j_{\mu 4}/\ka$ actually leads to the
introduction of fifth coordinates of position and momentum $x_4,p_4$ satisfying
\be
X_\mu=\f{1}{\ka}(x_\mu p_4-x_4 p_\mu).
\ee
A straightforward matching of this condition with the definition of $X_\mu$ gives:
\be
x_4=\f{1}{\ka}f\vphi,\qquad p_4=\ka f.
\ee
Taking into account the values of $f$ and $\vphi$, these new coordinates match the 5d
representation of DSR in the Snyder basis \cite{jerzy,obs}. Moreover, the present ``effective
coordinates" point of view gives a natural physical interpretation of the DSR fifth coordinates $x_4$
and $p_4$, as the shift and rescaling between the original (fundamental) space-time coordinate and
the effective (measured) coordinates.

We introduce the rescaled momentum $P_\mu=p_\mu/f$ dual to $X$. It is such that the representation
of the Lorentz transformations is not modified, $j_\mn=X_{[\mu}P_{\nu]}$. We also compute the new
canonical bracket:
\bes
\{X_\mu, P_\nu\}&=&
\eta_\mn-\f{P_\mu P_\nu}{\ka^2}\left[(f^2)'+\vphi'\left(1-2\f{p^2}{\ka^2}\f{f'}{f}\right)\right] \nonumber\\
%&=&
%\eta_\mn-\f{P_\mu
%P_\nu}{\ka^2}\left[\eps+\alpha\f{\eps'-\eps\f{p^2}{\ka^2}}{\eps'+\eps\f{p^2}{\ka^2}}\right] \\
&=&
\eta_\mn-\f{P_\mu
P_\nu}{\ka^2}\left[\eps+\alpha-2\alpha\eps\f{P^2}{\ka^2}\right],
\ees
where we have assumed $\eps'\ne 0$. The case $\eps'=0$ is somewhat pathological since it means that
$P^2$ does not vary and is always normalized to $\eps\ka^2$. We therefore exclude this case and
restrict our analysis to $\eps'=\pm1$.

\medskip

Let us look at the physical interpretation of these effective coordinates. We focus on one case
$\eps=\eps'=1$, but everything can be easily transposed to the other cases. The 5d structure is
then $\so(4,1)$. Looking at the momentum variables, we have:
\be
P_\mu=\f{p_\mu}{\sqrt{1+\f{p^2}{\ka^2}}},\qquad p_\mu=\f{P_\mu}{\sqrt{1-\f{P^2}{\ka^2}}}
\ee
%$$
%\left(1+\f{p^2}{\ka^2}\right)\left(1-\f{P^2}{\ka^2}\right)=1
%$$
In the $p$ variables, we have a truncation of the phase space $p^2\ge -\ka^2$. On the other side,
we have a restriction on the mass in the $P$ variables, $P^2\le \ka^2$. This maximal cut-off of the
momentum $P$ is dual to a minimal length scale $\hbar/\ka$ in coordinate space $X$.

The important point is that if we effectively measure the variables $P_\mu$ as momentum the the
conservation laws will look deformed. Indeed, considering the addition law for momenta, the
addition $p_\mu\equiv p_\mu^{(1)}+p_\mu^{(2)}$ will become non-trivial expressed in the $P$
variables:
\be
\f{P_\mu}{\sqrt{1-\f{P^2}{\ka^2}}}\equiv \f{P^{(1)}_\mu}{\sqrt{1-\f{(P^{(1)})^2}{\ka^2}}} +
\f{P^{(2)}_\mu}{\sqrt{1-\f{(P^{(2)})^2}{\ka^2}}}.
\ee
This defines a deformed addition for effective momenta $P_\mu=P_\mu^{(1)}\oplus P_\mu^{(2)}\ne
P_\mu^{(1)}+P_\mu^{(2)}$  although the underlying physics has not been modified. From this point of
view, the $P$'s are not the fundamental variables, however they are the variables that we
effectively have access to through direct measurements. Let us point out that this deformed
addition is nevertheless still commutative. %This fits with some formulations of DSR but clashes
%with the standard $\ka$-Poincar\'e approach.
%\footnotetext{This approach is therefore different than the usual approach where the $p$ would be the measured coordinates, whereas the $P$ are the platonic variables}

This perspective also allows to use different deformation mass scales for the different systems,
$\ka^{(1)},\ka^{(2)},\ka$. $\ka$ could a universal scale, or depend on the space-time curvature, or
be related to the physical properties of the system, or even depend on the observer and the choice
of measurements \cite{ESR,daniele}.

Then does there exist a natural choice for $\ka$ in term of $\ka^{(1)}$ and $\ka^{(2)}$? One
possibility is the existence of a fundamental mass scale, such as the Planck mass $M_P$. Then we
could take $\ka=\ka^{(1)}=\ka^{(2)}=M_P$. This is the traditional choice in DSR. Nevertheless, the
five-dimensional structure suggests a different approach. Indeed, we can use the new fifth
component of the momentum and postulate a new conservation law:
\be
p_4=p_4^{(1)}+p_4^{(2)}.
\ee
For small energies, when the $p^2$'s are small compared to the $\ka$'s, the leading order of this
fifth addition law reduces to $\ka\approx \ka^{(1)}+\ka^{(2)}$. This is exactly the 5d point of
view on DSR, which we actually proposed to call {\it Extended Special Relativity}  to
emphasize the difference with the standard formulation of DSR \cite{ESR}.

\medskip

%%%%
%\section{Conclusion}
%%%%

We have considered  a shift and a rescaling (both Lorentz covariant and momentum dependent)  of the
space-time coordinates, which shows how the notion of minimum length can appear at an effective
level within special relativity. This formalism can naturally be recast as a five-dimensional
framework and related to Snyder's approach for a Lorentz invariant non-commutative space-time. Our
point of view then provides the missing physical interpretation of the extra 5d coordinates $(x_4,
p_4)$: they precisely encode the information about the shift and the rescaling. As a consequence,
the deformed addition of (effective) momenta, which is commutative, also encodes the natural
rescaling of the deformation mass scale
%when considering many particles,
avoiding therefore the ``soccer-ball problem" often met in theories with a minimal length such as
DSR.

%\section*{Acknowledgements}

%%%%%%%%%%%%%%%%%

\end{document}